\documentclass[twocolumn,showpacs,preprintnumbers,amsmath,amssymp]{revtex4}
\usepackage{graphicx}
\usepackage{bm}
 \newcommand{\beq}{\begin{equation}}
 \newcommand{\eeq}{\end{equation}}
 \newcommand{\ber}{\begin{eqnarray}}
 \newcommand{\eer}{\end{eqnarray}}
 \newcounter{saveeqn}%
\newcommand{\alpheqn}{\setcounter{saveeqn}{\value{equation}}%
\stepcounter{saveeqn}\setcounter{equation}{0}%
\renewcommand{\theequation}
             {\mbox{\arabic{saveeqn}-\alph{equation}}}}%
\newcommand{\reseteqn}{\setcounter{equation}{\value{saveeqn}}%
\renewcommand{\theequation}{\arabic{equation}}}%
\begin{document}

\title{Phase Transition between `A' and `B' forms of DNA: A Free Energy
 Perspective}

\author{Devashish Sanyal}
\email{deva@iopb.res.in}
\affiliation{ Theoretical Condensed Matter \\
Institute of Physics \\
 Bhubaneswar 751005, India 
 }
\date{\today}
\begin{abstract}
 We study the structural transition from `B' form of DNA to `A' 
form of DNA
 using group
 theoretic
 methods.  The transition is not of the order-disorder type and hence to
 construct a Landau kind of theory for the transition we  define a higher
 symmetry and relevant order parameters.
 We also discuss  the issue of all the conformations, observed experimentally
 during the course of transition, being fundamentally different or not.
\end{abstract}
\maketitle
\section{Introduction}

 DNA can exist in many conformations, the most stable among them being
 `A', `B'  and `Z' forms.`A', `B' forms of DNA  form right-handed double
 helix while `Z'
 DNA forms left-handed double helix with varying number of turns  per
 rotation. In this report we examine the transition between `B' and
 `A' forms theoretically. `B'
 DNA is the most commonly occurring DNA existing in highly humid conditions.
 It may be transformed to the `A' form by  methylation. The transition from
 `B' form to
 `A' form was observed during the first 
 x-ray studies by Franklin and Goslin[1]
 who characterized the change of the helix
 structure as ` a substantial re-arrangement
 of the molecule'. Experiments[2,3] have been performed to observe the 
 transition from `B' form to `A' form in a more detailed manner and this
 report is an attempt to explain the fundamental features of the experiments. 
  Recent experiments[4] to
 study the dynamics of the
 transition from `B' to `A ' observe
 that  the helical states are clearly
 separated from one another in terms of  activation energy with the time
 scale of the dynamics of transition being
 of the order of ms.
  In the following paragraphs
 we will try to address the question of whether the  transition
 from `B' form to `A' form is a phase transition as defined
 in statistical physics. The study of the
 `B' to `A'
 transition from
 the first principles by including the interaction between the various
 bases and their interaction with the environment can be very challenging.
 The Landau analysis may provide a sophisticated method to study the
 phase transition.
\section{ Symmetry of `A' and `B' Forms}
 The  Landau theory for the study of  phase transition is considered as
 one of the important achievements of Landau. In Landau theory  the
 free energy is written in terms of the order parameter which  in
 turn depends on the symmetry
 of the  phase having the higher symmetry through
 one of its irreducible representations. The free energy remains invariant
 under the symmetry operations of the higher symmetry. Phase
 transitions  involving multiferroics [5],complicated structures
  such as  intercalation compounds[6]
 and antiferromagnets[7]  have been studied  using the Landau  theory .
 In this report the Landau theory will be used to study the phase transition
 between `A' and `B' forms of DNA.
     Experimental studies , among other techniques, use  crystallography
 to study the structure of  DNA. The various conformations take on different 
 crystallographic structures[2,3] corresponding to different space groups.
 The space groups depend on the detailed symmetry of the molecule, length
 of the sequence etc. The use of space groups may not be the most
 reliable method to study
 the transition theoretically. After all, the DNA exists as a single macro molecule
 and we would prefer to treat it like that.\\
In the present approach the DNA is considered to be double helix with
 Watson-Crick bonding between the two strands.
 We would not go into
 the detailed coordinates of the atoms  in the two forms but would like
 to explain the transition from the point of view of change in symmetry
 resulting in the difference between `A' and `B' forms.
With this objective in mind, we consider the DNA macro molecule
 to be infinitely long along the axis. Hence , it can be looked upon as
 a transition where the line group changes[8].The view along the
 axis of `A' and `B' DNA is shown in Fig1[9].
 \begin{figure}[h]
\includegraphics[angle=0,scale=1.2]{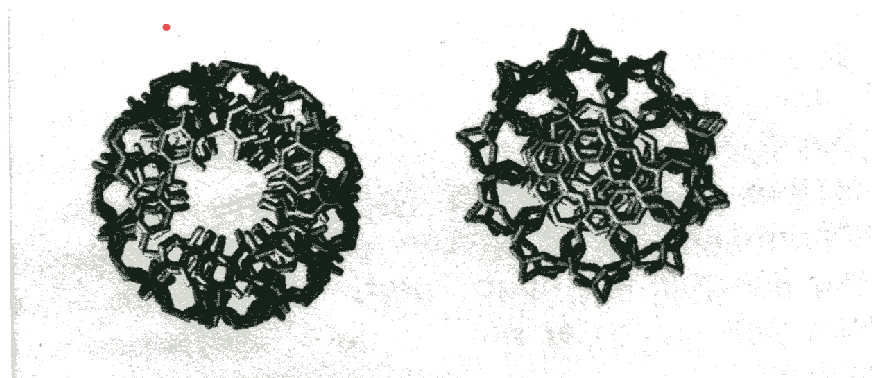}\caption{\small\texttt {Axial
 view of `A' and `B' forms}}
\end{figure} 
Apart from a   pure translational
 symmetry `a' along the axis, `B' form has the screw symmetry $10_1$. The
 rotation part of the group is cyclic with order 10($C_{10}$)[9]. `A' form
 of DNA has screw symmetry $11_1$. The rotation part of the
 group is cyclic with order 11($C_{11}$). In the subsequent analysis we
 employ group theoretic arguments to construct the order parameter in 
 order to write down the free energy. The irreducible representation for
 the rotation part, say of order $\alpha$, may be deduced from the  relation
$z^{\alpha}=1$ as the group is cyclic. So $z=e^{\frac{i2\pi n}{\alpha}}$ where
 $n=0,1...,\alpha -1$. Since
 the groups($C_{10}$,$C_{11}$) are abelian, all the representations
 are one dimensional . $D^{n}$ represents the $n^{th}
$ irreducible representation.

 The basis for the representations is given by $e^{-in\phi}$ where
 $n=0,1,..9$ for $C_{10}$ and $n=0,1,2..10$ for $C_{11}$. $R(\Delta\phi)$
 be an operator that rotates the coordinate system by $\Delta\phi$.
 If $\Delta\phi=\frac{2\pi}{10}$(say), then $R(\Delta\phi)e^{-in\phi}=
 e^{\frac{i2n\pi}{10}}e^{-in\phi}$, n being the index of representation.
 Thus $e^{\frac{i2n\pi}{10}}$ is a representation of\emph
 rotation by $2\pi/10$ with 
 index $n$. We can thus calculate the representations of $C_{10}$
 and $C_{11}$ corresponding to the various irreducible representations.
 The irreducible 
 representation of the line group is given by $e^{-in\phi}e^{ikz}$,
 $e^{ikz}$ being the  basis for the representation of translation
 along  the  DNA axis. The representation  is labeled by the pair
 $(n,k)$.
\section{ Construction of the order parameter}
 The order parameter, which is used to describe order-disorder transition,
 is the expectation value of an operator which does
 not remain invariant under the full symmetry group, taking the value $0$
 in the disordered phase and a finite value in the ordered phase.It,
 though, remains invariant under a subgroup of transformations.
 In order
 to introduce the  order parameters in the problem we need to  define
 disordered phase( high symmetry phase). We consider $10_1$(`B' DNA) and
 $11_1$(`A' DNA) as the ordered
 phases in comparison to the disordered phase(high symmetry phase) given by the 
 symmetry $(10\times11\times m)_1$, m being some integer.
 This is
 so as  they are subgroups of the high symmetry phase. We take $m=1$  in this
 case for the sake of convenience.  
Let $R(\frac{2\pi q}{10})T(\frac{qa}{10})$
 $(q=0...9)$ be a symmetry operation of `B' DNA with R  standing
 for the rotation and T standing for the translation.
\ber
R(\frac{2\pi q}{10})T(\frac{qa}{10})e^{-in\phi}e^{ikz}=
 e^{\frac{i2\pi nq}{10}}e^{\frac{ikqa}{10}}e^{-in\phi}e^{ikz}
\eer
We want $ e^{\frac{i2\pi nq}{10}}e^{\frac{ikqa}{10}}$ to be unity
 since the order parameter transforms as some irreducible
 representation and remains invariant under a subgroup of
 transformation in the ordered phase. Hence
 $\frac{2\pi n}{10} + \frac{ka}{10} = 2\pi A$ , A being an integer. It leads to
\ber 
k=\frac{2\pi(10A-n)}{a}. 
\eer
One may repeat the same arguments for `A' DNA to obtain,
\ber
k=\frac{2\pi(11A-n)}{a}.
\eer
\begin{figure}[h]
\includegraphics[ angle=0,scale=0.5]{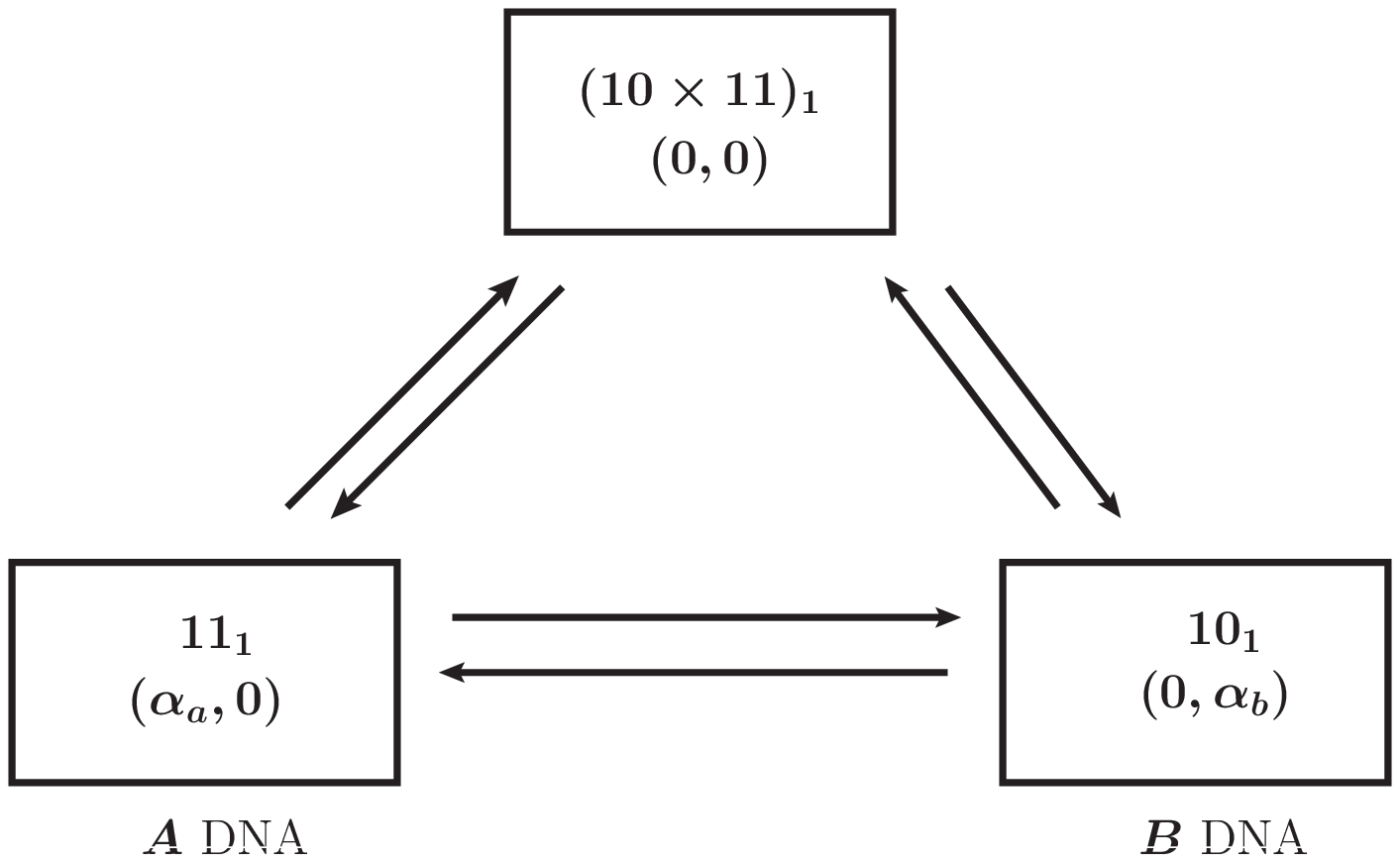}\caption{\small\texttt{
 Schematic diagram representing the Landau  theory for transition
 between `A' and `B' forms of DNA}}
\end{figure}

 We note that $n=0,..109$ since the higher symmetry is denoted by
 $(10\times11)_1$.  We take $n=1$, $A=1$ for convenience.
 For `B' DNA $k=\frac{18\pi}
{a}$ and for `A' form $k=\frac{20\pi}{a}$ though there could be different 
 values of $k$ corresponding to different values of $n$ and $A$.
 We would work in the 
 space of basis $(k_1, n_1)$ and $(k_2, n_2)$ where $k_1=\frac{20\pi}{a}$
 ,$n_1=1$ and $k_2=\frac{18\pi}{a}$,$n_2=1$. There is no mixing between
$(k_1, n_1)$  and  $(k_2, n_2)$  , so the system is defined by
  two order parameters $\alpha_a$ and $\alpha_b$
 transforming respectively as the representations determined by the above
 basis. $(\alpha_a,0)$ is
 invariant under the symmetry operations of $11_1$ $-$ $\alpha_a$ remains
 invariant due to (3) and $ 0 $ remains so under the  operation $-$ thereby
 representing  the `A' DNA phase.
 phase. For similar reasons $(0,\alpha_b)$ being invariant 
under the operations of $10_1$ represents
 the `B' DNA phase . $(0,0)$ being invariant under the larger
 symmetry group represents
 the melt
 or the higher symmetry The picture  of the phase transition is presented
 in Fig2.
  We try to establish the transition between $B$ DNA and $A$ DNA by 
  looking at the transitions,
  possibly fictitious, between the disordered phase and the ordered
 phases.
 We consider $\alpha_a$ and $\alpha_b$ to be  complex
  quantities. Having constructed the two order parameters 
 we may now attempt to write down the free energy of the system.
 It may be seen that under a general operation of $(10\times 11)_{1}$
 $\alpha_a$ and $\alpha_b$ picks up a phase. Hence a  very
 general form of the free energy in the form of a polynomial
 in $\alpha$ and $\alpha^{\ast}$ until the 
 quartic order, invariant under the operation
 of the higher symmetry group, may be written down.
\ber
\lefteqn{f=\frac{1}{2} r(\mid{\alpha_a}\mid ^2 + \mid{\alpha_b}\mid ^2)
 -\frac{1}{2} g
(\mid{\alpha_a}\mid ^2 - \mid{\alpha_b}\mid ^2)}\nonumber\\ &&
+ u_1\mid{\alpha_a}\mid ^4 + u_2\mid{\alpha_b}\mid ^4 +
 2u_{22}\mid{\alpha_a}\mid ^2\mid{\alpha_b}\mid ^2
\eer
 where $u_1$,$u_2$ $>$ $0$.
It may be seen here that the  coefficients of $\mid{\alpha_a}\mid ^2,
\mid{\alpha_b}\mid ^2$
 have been split among $r$ and $g$ as they separately carry  different
 information.
 The transition from  `B' DNA to `A' DNA is carried out through
 the addition of a chemical agent($c$). In the above equation
 we put $r\propto
 (T-T^{\star})$ and $g\propto (c-c^{\star})$, $T^{\star}$
 and $c^{\star}$(chemical conc) are some constants and $T$, not necessarily
 the temperature, may not be a  realizable parameter. 
$u_{22}$  is the term
 that is responsible for the coexistence of phases. The reason for this
 would be clear when we discuss the solutions to (4). 
  The minimization of the free energy gives the following
   combinations for the possible values of $\alpha_a$ and
 $\alpha_b$.
\alpheqn
\beq
\mid\alpha_a\mid=0,\mid\alpha_b\mid=0
\eeq
\beq
\mid\alpha_a=0\mid,\mid\alpha_b\mid\neq 0
\eeq
\beq
\mid\alpha_a\mid\neq 0,\mid\alpha_b\mid=0
\eeq
\beq
\mid\alpha_a\mid\neq 0,\mid\alpha_b\mid\neq 0
\eeq
\reseteqn
 We first
 consider
 the case where $u_{22}>0$. A complete analysis involving the stability of the
 solutions in the $r-g$ plane is shown below[10]. Fig3 corresponds to the
 case $u_1 u_2 - {u_{22}}^2 <0$ and Fig4 corresponds to the case 
  $u_1 u_2 - {u_{22}}^2 >0$
\begin{figure}[h]
\includegraphics[ angle=0,scale=0.8]{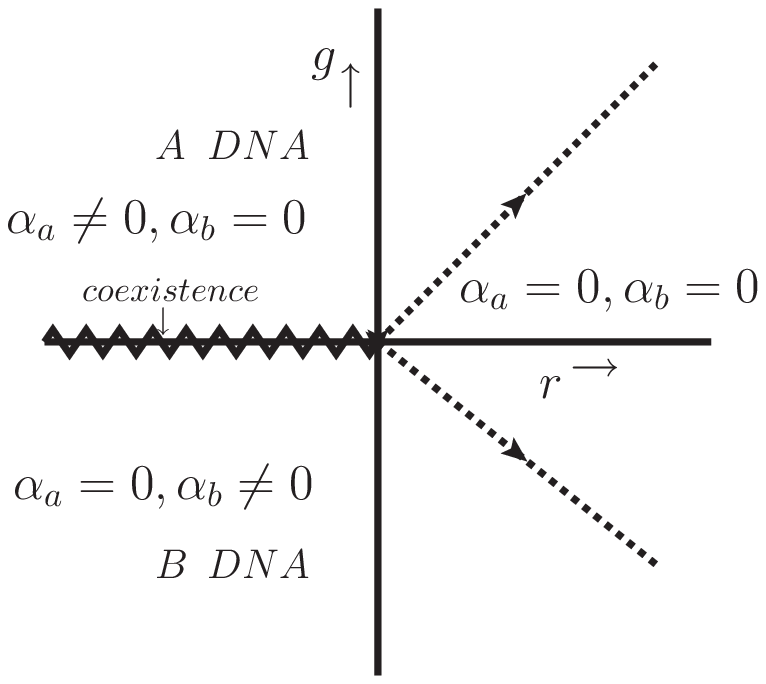}\caption{\small\texttt{
 Phase diagram corresponding to the free energy given by
 eqn(4) in the $r-g$ plane for
$u_1 u_2 - {u_{22}}^2<0$.  This diagram corresponds to the transition between
 `A' and `B' forms.}}
\end{figure}
\begin{figure}[h]
\includegraphics[angle=0,scale=0.8]{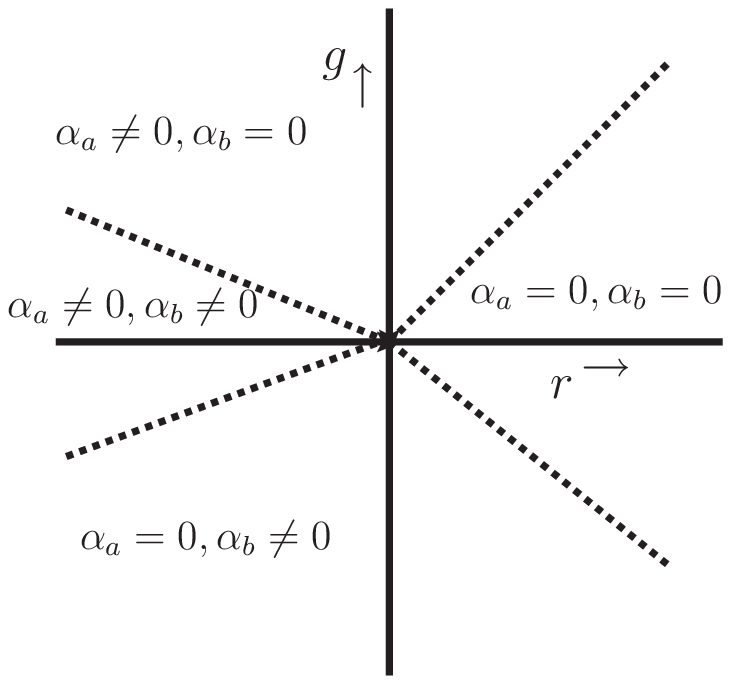}\caption{\small\texttt{
 Phase diagram corresponding to the free energy given by eqn(4) in the 
 $r-g$ plane for
$u_1 u_2 - {u_{22}}^2 >0$.}}
\end{figure} 
  The dotted lines in the  figures denote second order transition
 while the wiggly line denotes first order transition line.
 Being on the wiggly line signifies the state of
 coexistence as shown in Fig3. In Fig3  we observe that
 the disordered phase
 ($\mid\alpha_a=0\mid,\mid\alpha_b=0\mid$) is separated from the ordered
 phase(`B' form
 and `A' form) by the second order lines while the ordered phases themselves
 are separated by first order line. The second order and the first order
 lines meet at the bi critical point. In Fig4 we observe the  emergence
 of another ordered phase($\mid\alpha_a\mid\neq 0,\mid\alpha_b\mid\neq 0$)
 separated
 from
 the  other two ordered phases by  second order lines.
 If $u_{22}$ is sufficiently positive so that the condition
 for Fig3 is satisfied, we have coexistence, otherwise not.
 Considering $u_{22}\rightarrow 0$, the condition for Fig4 is
 easily satisfied and hence no phase coexistence as should be the 
 case.  For $u_1>0$,$u_2>0$ and $u_{22}>0$ we get consistent solutions
 for $(4)$. But if $u_{22}<0$,  it can be shown that the requirement
 of the stability of the solutions gives rise to inconsistent picture.
 This result is entirely an analytical outcome of $(4)$ and is independent
 of DNA.
   We have  used group theoretic arguments 
  to  construct the order parameters  and write the
 free energy expression. The free energy gives certain phase diagrams which
 can be explained  in the light of experimental results. It may be noted here 
 that any transition taking us from the group $p_1$ to $q_1$  may be 
 represented by a similar theory.
\section{Interpretation of the Order Parameter}
 Having constructed the order parameters  we would now
 like to identify them. In condensed matter systems, we may have multicomponent
 order parameter determined by the possible irreducible representation of the
 point group of the relevant vector $\overrightarrow{k}$ in the reciprocal
 lattice as
 well as
 the star of $\overrightarrow{k}$. In the case of antiferromagnets[6] we may
, for example,
 have the magnetization perpendicular to the antiferromagnetically coupled
 ferromagnetic planes or situated in the plane. For the intercalation
 compounds, the components of the order parameter are in a way related to the
 Fourier components of the density  of the alkali
 atoms corresponding to the stars
 of $\overrightarrow{k}$. In the
 present case,however,
 we are dealing with two single component order parameters for
 a system whose structure does not fall under the category of Bravais lattice.
 Hence the above prescription would not be followed  while  identifying the
 order parameter. Being a structural phase transition,
 we would  relate the order parameters to densities. For the double helix
 DNA, $11_1$, $10_1$ or $(10\times 11)_1$ symmetry uniquely defines the sites
 on the strands.
  $\rho_a$ 
 denotes the densities of sites satisfying $11_1$ symmetry and $\rho_b$ 
 the density of sites satisfying $10_1$. We identify the modulus
 of the order parameters
 as related to $\rho_a - \rho_b$ and $\rho_b-\rho_b=0$
 if $\rho_a >\rho_b$ and
 vice versa. In the 'A' phase $\rho_b=0$ and in the `B' phase $\rho_a=0$.
In the case of $\rho_a = \rho_b$
(disordered phase), we have for both the order parameters 
 $\rho_a - \rho_b=0$. 
 With the above definition of the order parameters,
 we are able to identify the two ordered phases and the the disordered phase,
as depicted in Fig3. 

 Experiments that have been carried out to study the phase transition
  [2,3] report conformations in which both `A' type and `B' type
  features are there apart from only `A' type or `B' type
 features. We interpret
 this as coexistence of phases. This may suggest that the transformation
 is first-order in nature as depicted in Fig3,
 coexistence being a characteristic
 of first order transition.  The two phases coexist for the order parameters
 jump discontinuously at the first-order line. In first order phase
 transition when one goes from one phase to another there is  jump in
 the order parameter at the transition point . Similarly when one goes
 back to the same phase there is a jump. This only implies coexistence of
 phases at the transition point. 
Further, it may be noted that
 the DNA molecule, under experimentation in [2,3], being not a long molecule,
 has a finite region of coexistence in the r-g plane whereas Fig3 shows a
 line(wiggle) representing the coexistence valid for infinitely long 
 molecule.
$\mid\alpha_a\mid\neq 0$ , $\mid\alpha_b\mid\neq 0$ is the most
 ordered phase and in our problem should reflect a phase that has the
 lowest symmetry.But such a phase   which will have a lower symmetry
  than both $10_1$ and $11_1$ is not possible.
  This perhaps tells us that the 
 Fig3 represents the likely phase picture and Fig4 may be ruled out.
 \\
 If we refer to the experimental results of [2,3], it is mentioned that the
 transition from `B' DNA to `A' DNA takes place through 14 conformations.
 Further, the experimental data in  [2,3], where they have measured
 certain quantities like  tilt angle, axial distance etc do not
 actually tell us whether there has been a phase transition or not. But in the
  light of the above analysis, it may be concluded that
 the transition from `B' to `A' is a first order phase transition
 with  all the conformations
  being  not fundamentally different but belong to  different phases(`B' and
 `A') and the region of coexistence.  

\noindent
\vspace{0.5cm}

{\large \bf Acknowledgements}
\vskip0.2cm
     Criticism by the referees is gratefully acknowledged.

\end{document}